\documentclass[aps,prl,showpacs,amsmath,amssymb,amsfonts,superscriptaddress,twocolumn,lengthcheck]{revtex4-1}
\usepackage [latin1]{inputenc}
\usepackage{amsmath}
\usepackage{amsfonts}
\usepackage{graphicx}
\usepackage{epsfig}
\usepackage{color}
\usepackage[colorlinks,citecolor=blue]{hyperref}
\begin{document}

\title{Noise-Tolerant Optomechanical Entanglement via Synthetic Magnetism}

\author{Deng-Gao Lai}
\affiliation{Key Laboratory of Low-Dimensional Quantum Structures and Quantum Control of Ministry of Education, Key Laboratory for Matter Microstructure and Function of Hunan Province, Department of Physics and Synergetic Innovation Center for Quantum Effects and Applications, Hunan Normal University, Changsha 410081, China}
\affiliation{Theoretical Quantum Physics Laboratory, RIKEN Cluster for Pioneering Research, Wako-shi, Saitama 351-0198, Japan}

\author{Jie-Qiao Liao}
\email{jqliao@hunnu.edu.cn}
\affiliation{Key Laboratory of Low-Dimensional Quantum Structures and Quantum Control of Ministry of Education, Key Laboratory for Matter Microstructure and Function of Hunan Province, Department of Physics and Synergetic Innovation Center for Quantum Effects and Applications, Hunan Normal University, Changsha 410081, China}

\author{Adam Miranowicz}
\affiliation{Theoretical Quantum Physics Laboratory, RIKEN Cluster for Pioneering Research, Wako-shi, Saitama 351-0198, Japan}
\affiliation{Institute of Spintronics and Quantum Information, Faculty of Physics, Adam Mickiewicz University, 61-614 Pozna\'{n}, Poland}

\author{Franco Nori}
\affiliation{Theoretical Quantum Physics Laboratory, RIKEN Cluster for Pioneering Research, Wako-shi, Saitama 351-0198, Japan}
\affiliation{Physics Department, The University of Michigan, Ann Arbor, Michigan 48109-1040, USA}

\begin{abstract}
Entanglement of light and multiple vibrations is a key resource for multi-channel quantum information processing and memory. However, entanglement generation is generally suppressed, or even fully destroyed, by the dark-mode (DM) effect induced by the coupling of multiple degenerate or near-degenerate vibrational modes to a common optical mode. Here we propose how to generate optomechanical entanglement via \emph{DM breaking} induced by synthetic magnetism. We find that at nonzero temperature, light and vibrations are \emph{separable} in the DM-unbreaking regime but \emph{entangled} in the DM-breaking regime. Remarkably, the threshold thermal phonon number for preserving entanglement in our simulations has been observed to be up to \emph{three} orders of magnitude stronger than that in the DM-unbreaking regime. The application of the DM-breaking mechanism to optomechanical networks can make noise-tolerant entanglement networks feasible. These results are quite general and can initiate advances in quantum resources with immunity against both dark modes and thermal noise.
\end{abstract}

\maketitle

\emph{Introduction.}---Quantum entanglement~\cite{Horodecki2009RMP}, allowing for inseparable quantum correlations shared by distant parties, is a crucial resource for modern quantum technologies, including quantum metrology, communication, and computation~\cite{Duarte2021Press}. So far, efficient entanglement of photons with atoms~\cite{Raimond2001RMP,Volz2006PRL,Wilk2007Science,Yuan2008Nature,Matsukevich2005PRL,Sherson2006Nature,Qin2018PRL}, trapped ions~\cite{Leibfried2003RMP,Blinov2004Nature}, quantum dots~\cite{Gao2012Nature}, and superconducting qubits~\cite{You2011Nature,Gu2017PR,Xiang2013RMP} has been demonstrated in both microscopic- and macroscopic-scale devices~\cite{Jost2009Nature,Hensen2015Nature}. These entangled states have been used to connect remote long-term memory nodes in distributed quantum networks~\cite{Kimble2008Nature,Chou2005Nature,Moehring2007Nature,Safavi-Naeini2019Optica}.

The cavity optomechanical system is an elegant candidate for implementing quantum information carriers and memory~\cite{Kippenberg2008Science,Meystre2013AP,Aspelmeyer2014RMP,Bowen2015}. Owing to the remarkable progress in ground-state cooling~\cite{Wilson-Rae2007PRL,Marquardt2007PRL,Chan2011Nature,Teufel2011Nature} and single-phonon manipulation~\cite{Connell2010Nature,Chu20017Science,Gustafsson20014Science,Hong20017Science}, it has become an efficient platform for achieving quantum entanglement between two bosonic  modes~\cite{Vitali2007PRL,Tian2013PRL,Wang2013PRL,Vitali2007JPA,Mancini2002PRL,Hartmann2008PRL,Paternostro2007PRL,Huang2009NJP,Riedinger2016Nature,Palomaki2013Science,Barzanjeh2019Nature,Chen2020NC,Ho2018PRL,Jiao2020PRL,Yu2020Nature}.
In particular, macroscopic quantum entanglement involving two massive oscillators has recently been observed in optomechanical platforms~\cite{Riedinger2018Nature,Ockeloen-Korppi2018Nature,Kotler2021Science,Ockeloen-Korppi2021Science}. Practically, the applicability of modern quantum technologies in optomechanical networks ultimately requires quantum entanglement of light and many vibrations~\cite{Armstrong2012Nc,McCutcheon2016Nc,Cai2017Nc,Wengerowsky2018Nature}. Realization of large-scale photon-phonon entanglement, however, remains an outstanding challenge due to the suppression from the dark-mode effect~\cite{Scully1997QO,Agarwal2013QO,Lake2020Nc,Dong2012Science,Shkarin2014PRL,Sommer2019PRL,Massel2012Nc,Genes2008NJP,Ockeloen-Korppi2019PRA,Kuzyk2017PRA,Lai2020PRARC} induced by the coupling of an optical mode to multiple degenerate or near-degenerate vibrational modes~\cite{Shkarin2014PRL,Sommer2019PRL,Massel2012Nc,Genes2008NJP,Ockeloen-Korppi2019PRA,Kuzyk2017PRA,Lai2020PRARC} which could be used as multiple nodes in quantum networks~\cite{Kimble2008Nature}.

In this Letter, we propose to generate light-vibration entanglement by breaking the dark mode via synthetic magnetism, and reveal its \emph{counterintuitive robustness} to thermal noise. By introducing a loop-coupled structure, formed by light-vibration couplings and phase-dependent phonon-hopping interactions, a synthetic gauge field is induced and it breaks dark modes. Note that the realization of a reconfigurable synthetic gauge field has recently been reported in phase-dependent loop-coupled optomechanical platforms~\cite{Schmidt2015optica,Shen2016NP,Ruesink2016NC,Fang2017NP,Bernier2017NC,Shen2018NC,Ruesink2018NC,Mathew2018arXiv,Chen2021PRL}.
We find that in the dark-mode-unbreaking (DMU) regime, light-vibration entanglement is destroyed by thermal noise concealed in the dark modes; while in the dark-mode-breaking (DMB) regime, optomechanical entanglement is generated via synthetic magnetism. Surprisingly, in the DMB regime, the threshold phonon number for preserving entanglement has been observed to be up to \emph{three} orders of magnitude stronger than that in the DMU regime. Our work describes a general mechanism, and it can provide the means to engineer and protect fragile quantum resources from thermal noises and dark modes, and pave a way towards noise-tolerant quantum networks~\cite{Kimble2008Nature,Wengerowsky2018Nature}.


\begin{figure}[tbp]
\center
\includegraphics[width=0.45 \textwidth]{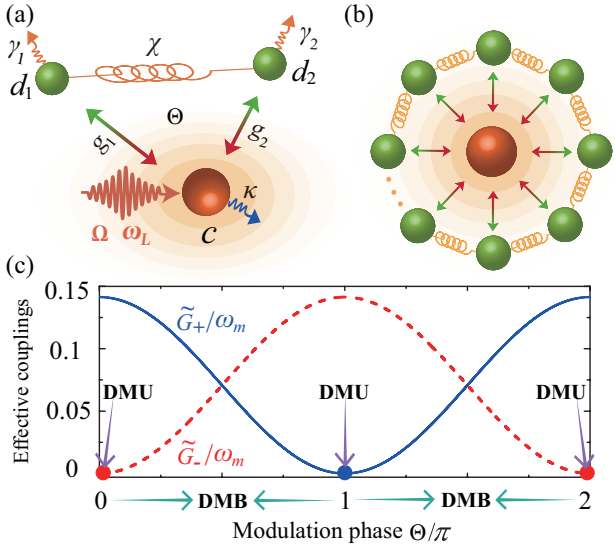}
\caption{(a) Loop-coupled optomechanical system consisting of two vibrational modes $d_{j=1,2}$ (with decay rates $\gamma_{j}$) coupled to a common optical mode $c$ (with decay rate $\kappa$) via optomechanical interactions (with strengths $g_{j}$). The two vibrations are coupled to each other through a phase-dependent phonon-exchange coupling ($\chi$ and $\Theta$). (b) An optomechanical network consisting of an optical mode coupled to $N$ vibrational modes. (c) Effective coupling strengths $\tilde{G}_{\pm}$ versus $\Theta$. By tuning $\Theta\neq n\pi$ for an integer $n$, the dark mode can be broken ($\tilde{G}_{\pm}\neq0$). The solid disks denote the DMU regime, and the remaining areas correspond to the DMB regime. Here we choose $\omega_{m}$ as the frequency scale and set $\omega_{j}/\omega_{m}=1$ and $G_{j}/\omega_{m}=\chi/\omega_{m}=0.1$.}
\label{Figmodel}
\end{figure}

\emph{System and dark-mode control.}---We consider a loop-coupled optomechanical system consisting of an optical mode and two vibrational modes [see Fig.~\ref{Figmodel}(a)]. A driving field, with frequency $\omega_{L}$ and amplitude $|\Omega|=\sqrt{2\kappa \mathcal{P}_{L}/(\hbar\omega_{L})}$ (with laser power $\mathcal{P}_{L}$ and cavity-field decay rate $\kappa$), is applied to the cavity field. In a rotating frame defined by the unitary transformation operator $\exp(-i\omega_{L}c^{\dagger}ct)$, the system Hamiltonian reads ($\hbar =1$)
\begin{eqnarray}
\mathcal{H}_{I}&=&\Delta_{c}c^{\dagger}c+\sum_{j=1,2}[\omega_{j}d_{j}^{\dagger }d_{j}+g_{j}c^{\dagger}c(d_{j}+d_{j}^{\dagger})]\notag \\
&&+(\Omega c+\Omega^{\ast}c^{\dagger})+\mathcal{H}_{\chi},\notag \\
\mathcal{H}_{\chi}&=&\chi(e^{i\Theta}d_{1}^{\dagger}d_{2}+e^{-i\Theta}d_{2}^{\dagger}d_{1}),\label{eq1iniH}
\end{eqnarray}
where $c^{\dagger}$ ($c$) and $d^{\dagger}_{j}$ ($d_{j}$) are the creation (annihilation) operators of the cavity-field mode (with resonance frequency $\omega_{c}$) and the $j$th vibrational mode (with resonance frequency $\omega_{j}$), respectively. The $g_{j}$ terms describe optomechanical interactions between the cavity-field mode and the two vibratinal modes, with $g_{j}$ being the single-photon optomechanical-coupling strength. The $\Omega$ term denotes the cavity-field driving with detuning $\Delta_{c}=\omega_{c}-\omega_{L}$. The $\mathcal{H}_{\chi}$ term depicts the phase-dependent phonon-hopping interaction (with coupling strength $\chi$ and modulation phase $\Theta$), which is introduced to create synthetic gauge fields and control the dark-mode effect~\cite{seeSM}. 

To demonstrate the dark-mode effect, we expand the operators $o\in$\{$c$, $d_{j}$, $c^{\dagger}$, $d^{\dagger}_{j}$\} as a sum of their steady-state average values and fluctuations, i.e., $o=\langle o\rangle_{\text{ss}}+\delta o$. Then we obtain the linearized Hamiltonian in the rotating-wave approximation (RWA) as: $\mathcal{H}_{\text{RWA}}=\Delta \delta c^{\dagger}\delta c+\sum_{j=1,2}[\omega_{j}\delta d_{j}^{\dagger}\delta d_{j}+G_{j}(\delta c\delta d_{j}^{\dagger}+\mathrm{H.c.})]+\chi(e^{i\Theta}\delta d_{1}^{\dagger}\delta d_{2}+\mathrm{H.c.})$,
where $\Delta$ is the normalized driving detuning~\cite{seeSM} and $G_{j=1,2}=g_{j}\langle c\rangle_{\text{ss}}$ are the linearized optomechanical-coupling strengths. Here $\langle c\rangle_{\text{ss}}=-i\Omega^{*}/(\kappa+i\Delta)$ is assumed to be real by choosing a proper $\Omega$. Note that the RWA of the light-vibration interaction is performed only in the demonstration of the dark-mode effect and its breaking. In the derivation of entanglement measures and the numerical simulations, we consider both the beam-splitter-type and two-mode-squeezing-type optomechanical interactions.

When the synthetic magnetism is absent (i.e., $\chi=0$) and $\omega_{1}=\omega_{2}$, the system possesses two hybrid mechanical modes: \emph{bright} ($\mathcal{D}_{+}$) and \emph{dark} ($\mathcal{D}_{-}$) modes defined by
\begin{align}
\mathcal{D}_{\pm}=&\;(G_{1(2)}\delta d_{1} \pm G_{2(1)}\delta d_{2})/\sqrt{G^{2}_{1}+G^{2}_{2}},\label{Dark}
\end{align}
which satisfy the bosonic commutation relation $[\mathcal{D}_{\pm},\mathcal{D}_{\pm}^{\dagger}]=1$. The dark mode $\mathcal{D}_{-}$, which decouples from the system and destroys all quantum resources, can be broken by employing the synthetic magnetism (i.e., $\chi\neq0$ and $\Theta\neq0$). To clarify this, we introduce two superposition-vibrational modes associated with the synthetic magnetism: $\tilde{\mathcal{D}}_{\pm}=\mathcal{F}\delta d_{1(2)}\mp e^{\pm i\Theta}\mathcal{K}\delta d_{2(1)}$, which satisfy the bosonic commutation relation $[\mathcal{\tilde{D}}_{\pm},\mathcal{\tilde{D}}_{\pm}^{\dagger}]=1$. Here $\mathcal{F} =\vert\delta\tilde{\omega}_{-}\vert/\sqrt{(\delta\tilde{\omega}_{-})^{2}+\chi^{2}}$ and $\mathcal{K} =\chi \mathcal{F}/\delta\tilde{\omega}_{-}$, with $\delta\tilde{\omega}_{-}=\tilde{\omega}_{-}-\omega_{1}$ and the redefined resonance frequencies $\tilde{\omega}_{\pm} =(\omega _{1}+\omega _{2}\pm \sqrt{(\omega_{1}-\omega_{2})^{2}+4\chi^{2}})/2$. The linearized Hamiltonian becomes
\begin{eqnarray}
\!\!\!\!\mathcal{H}_{\text{RWA}}\!\!=\!\!\Delta \delta c^{\dagger }\delta c+\!\!\sum_{l=\pm}[\tilde{\omega}_{l}\tilde{\mathcal{D}}_{l}^{\dagger}\tilde{\mathcal{D}}_{l}+(\tilde{G}_{l}\tilde{\mathcal{D}}_{l}\delta c^{\dagger}+\mathrm{H.c.})], \label{HRWA3}
\end{eqnarray}
where the effective coupling strengths are $\tilde{G}_{\pm}=\mathcal{F}G_{1(2)}\mp e^{\mp i\Theta}\mathcal{K}G_{2(1)}$.
In Fig.~\ref{Figmodel}(c), we show $\tilde{G}_{\pm}$ versus $\Theta$ when $\omega_{1}=\omega_{2}$ and $G_{1}=G_{2}$. We see that only when $\Theta=n\pi$ (i.e., the DMU regime), either $\tilde{\mathcal{D}}_{+}$ [for an odd $n$, $\tilde{G}_{+}=0$ (blue disks)] or $\tilde{\mathcal{D}}_{-}$ [for an even $n$, $\tilde{G}_{-}=0$ (red disks)] becomes a dark mode. Tuning $\Theta\neq n\pi$ (for an integer $n$, i.e., the DMB regime) leads to a counterintuitive coupling of the dark mode to the optical mode, which indicates dark-mode breaking. Physically, a reconfigurable synthetic gauge field is realized by modulating $\Theta$, which results in a flexible switch between the DMB and DMU regimes.

\emph{Langevin equations and their solutions.}---By defining the optical and mechanical quadratures $\delta X_{o}=(\delta o^{\dagger}+\delta o)/\sqrt{2}$ and $\delta Y_{o}=i(\delta o^{\dagger}-\delta o)/\sqrt{2}$, and the corresponding Hermitian input-noise operators $X^{\mathrm{in}}_{o}=(o_{\mathrm{in}}^{\dagger}+o_{\mathrm{in}})/\sqrt{2}$ and $ Y^{\mathrm{in}}_{o}=i(o_{\mathrm{in}}^{\dagger}-o_{\mathrm{in}})/\sqrt{2}$, we obtain the linearized Langevin equations as
$\mathbf{\dot{u}}(t)=\mathbf{Au}(t)+\mathbf{N}(t)$,
where we introduce the fluctuation operator vector $\mathbf{u}(t)=[\delta X_{d_{1}}, \delta Y_{d_{1}}, \allowbreak \delta X_{d_{2}}, \delta Y_{d_{2}},
\delta X_{c}, \delta Y_{c}]^{T}$, the noise operator vector $\mathbf{N}(t)=\sqrt{2}[\sqrt{\gamma _{1}}X_{d_{1}}^{\text{in}},\sqrt{\gamma_{1}}Y_{d_{1}}^{\text{in}},\sqrt{\gamma _{2}}X_{d_{2}}^{\text{in}}, \sqrt{\gamma _{2}}Y_{d_{2}}^{\text{in}},\sqrt{\kappa }X_{c}^{\text{in}},\sqrt{\kappa }Y_{c}^{\text{in}}]^{T}$, and the coefficient matrix
\begin{equation}
\mathbf{A}=\left(
\begin{array}{cccccc}
-\gamma _{1} & \omega _{1} & \chi _{+} & \chi _{-} & 0 & 0 \\
-\omega _{1} & -\gamma _{1} & -\chi _{-} & \chi _{+} & -2G_{1} & 0 \\
-\chi _{+} & \chi _{-} & -\gamma _{2} & \omega _{2} & 0 & 0 \\
-\chi _{-} & -\chi _{+} & -\omega _{2} & -\gamma _{2} & -2G_{2} & 0 \\
0 & 0 & 0 & 0 & -\kappa  & \Delta  \\
-2G_{1} & 0 & -2G_{2} & 0 & -\Delta  & -\kappa
\end{array}%
\right),
\end{equation}%
with $\chi _{+}=\chi \sin \Theta$ and $\chi _{-} =\chi \cos \Theta$.
The formal solution of the Langevin equation is given by $\mathbf{u}(t) =\mathbf{M}(t) \mathbf{u}(0)+\int_{0}^{t}\mathbf{M}(t-s)\mathbf{N}(s)ds$, where $\mathbf{M}(t)=\exp(\mathbf{A}t)$. Note that the parameters used in our simulations satisfy the stability conditions derived from the Routh-Hurwitz criterion~\cite{DeJesus1987PRA}.
The steady-state properties of the system can be inferred based on the steady-state covariance matrix $\mathbf{V}$, which is defined by the matrix elements
$\mathbf{V}_{kl}=[\langle \mathbf{u}_{k}(\infty) \mathbf{u}_{l}(\infty ) \rangle +\langle \mathbf{u}_{l}( \infty) \mathbf{u}_{k}(\infty )\rangle]/2$ for $k,l=1$-$6$. Under the stability conditions, the covariance matrix $\mathbf{V}$ fulfills the Lyapunov equation
$\mathbf{A}\mathbf{V}+\mathbf{V}\mathbf{A}^{T}=-\mathbf{Q}$~\cite{Vitali2007PRL},
where $\mathbf{Q}=\mathrm{diag} \{\gamma_{1}(2\bar{n}_{1}+1),\gamma_{1}(2\bar{n}_{1}+1),\gamma_{2}(2\bar{n}_{2}+1),\gamma_{2}(2\bar{n}_{2}+1),\kappa,\kappa\}$.

\emph{Generating bipartite entanglement and full tripartite inseparability via DMB.}---The logarithmic negativity $E_{\mathcal{N},j}$ and the \emph{minimum} residual contangle $E^{r|s|t}_{\tau}$, which can be used to quantify bipartite entanglement and full tripartite inseparability~\cite{Vidal2002PRA,Adesso2007JPA,Li2018PRL,Coffman2000PRA,Teh2014PRA}, are, respectively, defined as
\begin{subequations}
\label{23}
\begin{align}
E_{\mathcal{N},j}\equiv&\max [0,-\mathrm{ln}(2\zeta^{-}_{j})],\\
E^{r|s|t}_{\tau}\equiv& \min \limits_{(r,s,t)}[E^{r|(st)}_{\tau}-E^{r|s}_{\tau}-E^{r|t}_{\tau}].
\end{align}
\end{subequations}
Here $\zeta_{j}^{-}\equiv 2^{-1/2}\{\Sigma(\mathbf{V}^{'}_{j})-[\Sigma(\mathbf{V}^{'}_{j})^{2}-4 \mathrm{det} \mathbf{V}^{'}_{j}]^{1/2}\}^{1/2}$, with $\Sigma(\mathbf{V}^{'}_{j})\equiv\mathrm{det}\mathcal{A}_{j}+\mathrm{det}\mathcal{B}-2\mathrm{det}\mathcal{C}_{j}$, is the smallest eigenvalue
of the partial transpose of the reduced correlation matrix $\mathbf{V}^{'}_{j}=\left(\begin{array}{cc}
\mathcal{A}_{j} & \mathcal{C}_{j} \\
\mathcal{C}_{j}^{T} & \mathcal{B}\end{array}\right)$, which is obtained by removing in $\mathbf{V}$ the rows and columns of the uninteresting mode~\cite{seeSM}. In Eq.~(\ref{23}b), $r,s,t\in\{d_{1},d_{2},c\}$ denote all the permutations of the three mode indices. $E^{r|(st)}_{\tau}$ ($E^{r|s}_{\tau}$ or $E^{r|t}_{\tau}$) is the contangle of subsystems of $r$ and $st$ ($s$ or $t$), and it is defined as the squared logarithmic negativity~\cite{Adesso2007JPA,seeSM}. The residual contangle satisfies the monogamy of quantum entanglement $E^{r|(st)}_{\tau}\geq E^{r|s}_{\tau}+E^{r|t}_{\tau}$, which is based on the Coffman-Kundu-Wootters monogamy inequality~\cite{Coffman2000PRA}.
$E_{\mathcal{N},j}>0$ and $E^{r|s|t}_{\tau}>0$ mean, respectively, the emergence of bipartite optomechanical entanglement and full tripartite inseparability. The full inseparability is an important quantum resource and it is a necessary (but insufficient) condition for the presence of genuine multipartite entanglement~\cite{Teh2014PRA,Teh2022PRA,Armstrong2012Nc,Armstrong2015NP}.  In particular, this inseparability has recently been detected in experiments~\cite{Armstrong2012Nc,Armstrong2015NP}.

\begin{figure}[tbp]
\center
\includegraphics[width=0.45 \textwidth]{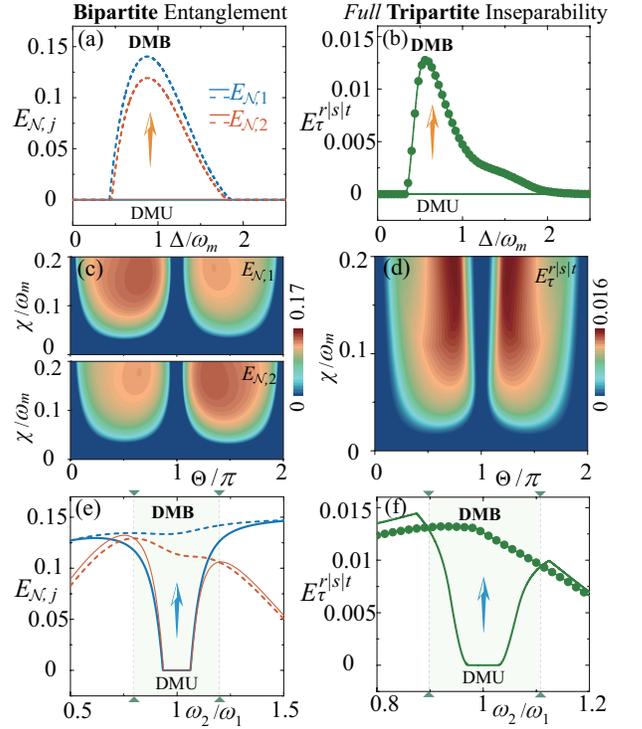}
\caption{(a) Bipartite entanglement measure $E_{\mathcal{N},j=1,2}$ and (b) full tripartite inseparability measure $E^{r|s|t}_{\tau}$ versus $\Delta/\omega_{m}$ in the DMU ($\chi=0$, horizontal solid lines) and DMB ($\chi=0.1\omega_{m}$ and $\Theta=\pi/2$, dashed curves and symbols) regimes. Here we set $\omega_{j}/\omega_{m}=1$ and choose $\omega_{m}$ as the frequency scale. (c) $E_{\mathcal{N},j}$ and (d) $E^{r|s|t}_{\tau}$ versus $\chi$ and $\Theta$ under the optimal drivings $\Delta=\omega_{m}$ for $E_{\mathcal{N},j}$ and $\Delta=0.6\omega_{m}$ for $E^{r|s|t}_{\tau}$. (e) $E_{\mathcal{N},j}$ and (f) $E^{r|s|t}_{\tau}$ versus $\omega_{2}/\omega_{1}$ in both DMU and DMB regimes, when $\omega_{1}/\omega_{m}=1$. Other parameters are $G_{j}/\omega_{m}=0.2$, $\gamma_{j}/\omega_{m}=10^{-5}$, $\kappa/\omega_{m}=0.2$, and $\bar{n}_{j}=100$.}
\label{detuning}
\end{figure}

We display in Figs.~\ref{detuning}(a,b) $E_{\mathcal{N},j}$ and $E^{r|s|t}_{\tau}$ versus the driving detuning $\Delta$, in both the DMU and DMB regimes. This demonstrates that light and vibrations are \emph{separable} in the DMU regime ($E_{\mathcal{N},j}=0$ and $E^{r|s|t}_{\tau}=0$, see the lower horizontal solid lines), but \emph{entangled} and \emph{full inseparability} in the DMB regime [$E_{\mathcal{N},1(2)}=0.14~(0.12)$ and $E^{r|s|t}_{\tau}=0.013$, see the upper dashed curves and symbols]. In the DMU regime, thermal phonons concealed in the dark mode cannot be extracted by the optomechanical cooling channel, and then quantum entanglement is completely destroyed by the residual thermal noise~\cite{Genes2008NJP,Sommer2019PRL}. In the DMB regime, a large entanglement is achieved around the red-sideband resonance ($\Delta\approx\omega_{m}$), corresponding to the optimal cooling. In addition, we have confirmed that in the blue-detuning regime, the introduced synthetic magnetism (i.e., the DMB mechanism) can enhance optomechanical entanglement~\cite{seeSM}. These results indicate that \emph{fragile quantum resources can be protected and engineered via dark-mode control}. 


In Figs.~\ref{detuning}(c,d), we plot $E_{\mathcal{N},j}$ and $E^{r|s|t}_{\tau}$ versus $\Theta$ and $\chi$. We can see that, at a finite value of $\chi$, $E_{\mathcal{N},j}$ and $E^{r|s|t}_{\tau}$ reach the peak value at $\Theta=\pi/2$ and $3\pi/2$, which are related to a strong quantum interference between two excitation-transport channels. In addition, both bipartite entanglement and full tripartite inseparability completely vanish, i.e., $E_{\mathcal{N},j}=0$ and $E^{r|s|t}_{\tau}=0$ at $\Theta=n\pi$, corresponding to the emergence of the dark mode. In particular, $E_{\mathcal{N},1}$ ($E_{\mathcal{N},2}$) is larger than $E_{\mathcal{N},2}$ ($E_{\mathcal{N},1}$) in the region $0<\Theta<\pi$ ($\pi<\Theta<2\pi$). This asymmetrical feature is caused by the modulation phase in the coupling loop, which corresponds to an effective synthetic magnetism~\cite{Fang2017NP,Bernier2017NC,Shen2018NC,Mathew2018arXiv,Chen2021PRL,Ruesink2018NC}. These findings mean that we can switch a multimode quantum device between separable and entangled states by tuning $\Theta$.


The dependence of bipartite entanglement and full tripartite inseparability on the dark-mode effect can also be seen by plotting $E_{\mathcal{N},j}$ and $E^{r|s|t}_{\tau}$ versus $\omega_{2}/\omega_{1}$ [see Figs.~\ref{detuning}(e,f)]. We find that in the DMU regime, there exists a disentanglement valley around the degeneracy point $\omega_{2}=\omega_{1}$, corresponding to the emergence of the dark-mode effect (see lower solid curves). However, in the DMB regime, the valley becomes smooth, which means that both bipartite entanglement and full tripartite inseparability fully survive owing to dark-mode breaking (see upper dashed curves and symbols). Physically, the width of the valley is determined by the spectral resolution in the three-mode system. Here, it is mainly determined by the cavity-field decay rate, because the decay rates of the mechanical modes are much smaller than that of the cavity field.

In addition, we see from Fig.~\ref{detuning}(e) that when $\omega_{1}\neq\omega_{2}$, the synthetic magnetism may slightly degrade the entanglement. This is because the dark-mode effect works within a finite frequency-detuning window, in which the breaking of the dark-mode effect dominates the enhancement of the entanglement generation. Out of this window, both the optomechanical and phonon-hopping interactions jointly contribute to the entanglement generation, and hence the constructive and destructive interferences caused by the two coupling channels in the loop-coupled system change optomechanical entanglement~\cite{seeSM}.

\begin{figure}[tbp]
\center
\includegraphics[width=0.45 \textwidth]{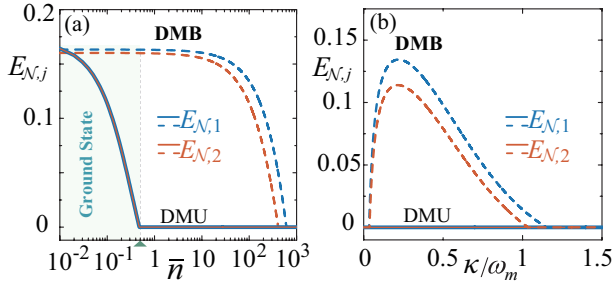}
\caption{(a) $E_{\mathcal{N},j=1,2}$ versus $\bar{n}_{j}=\bar{n}$ in the DMU (solid curves) and DMB (dashed curves) regimes. (b) $E_{\mathcal{N},j}$ versus $\kappa$ when $\bar{n}=100$. Here $\chi/\omega_{m}=0.1$ and $\Theta=\pi/2$, and other parameters are the same as those in Fig.~\ref{detuning}.}
\label{excitations}
\end{figure}

\emph{Noise-tolerant optomechanical entanglement}---The DMB entanglement provides a feasible way to create and protect fragile quantum resources against dark modes, and can enable the construction of noise-tolerant quantum devices. We can see from Fig.~\ref{excitations}(a) that, in the DMU regime, quantum entanglement emerges only when $\bar{n}\ll1$ (see solid curves), while in the DMB regime, it can persist for a threshold value near $\bar{n}\approx10^{3}$ (see dashed curves), which is \emph{three orders of magnitude} larger than that in the DMU regime.
In Fig.~\ref{excitations}(b), we plot $E_{\mathcal{N},j}$ versus $\kappa$ in both the DMB and DMU regimes. In the DMU regime, quantum entanglement is fully destroyed ($E_{\mathcal{N},j}=0$) by the thermal noise in the dark mode, and it is independent of $\kappa$ (see lower horizontal solid lines). However, in the DMB regime, light-vibration entanglement is generated owing to dark-mode breaking, and $E_{\mathcal{N},j}$ exist only in the resolved-sideband regime $\kappa/\omega_{m}<1$ (see upper dashed curves). The maximal entanglement is located at $\kappa\approx0.2\omega_{m}$, which is consistent with the typical deep-resolved-sideband conditions~\cite{Wilson-Rae2007PRL,Marquardt2007PRL,Chan2011Nature,Teufel2011Nature}. Electromechanical systems are excellent candidates for experimental demonstrations of noise-tolerant entanglement, because good cavities have been reported in this platform~\cite{seeSM}.

\begin{figure}[tbp]
\center
\includegraphics[width=0.45 \textwidth]{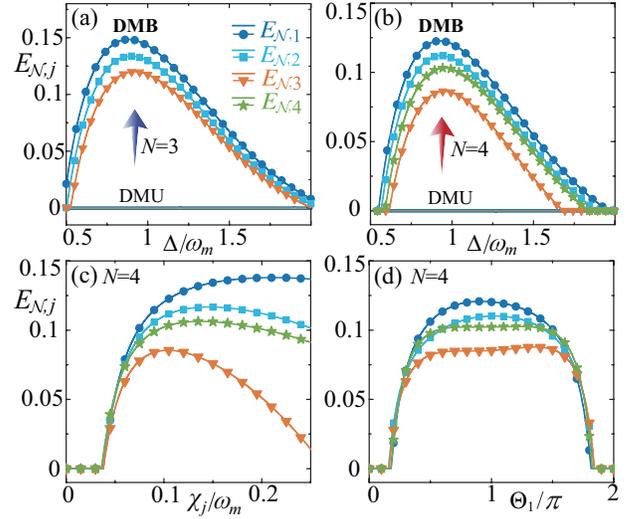}
\caption{(a,b) $E_{\mathcal{N},j}$ versus $\Delta$ in the DMU ($\chi_{j}=0$, the horizontal solid lines) and DMB ($\chi_{j}/\omega_{m}=0.1$ and $\Theta_{1}=\pi$, $\Theta_{j\in[2,N-1]}=0$, marked by symbols) regimes, for (a) $N=3$ and (b) $N=4$. (c) $E_{\mathcal{N},j}$ versus $\chi_{j}$ when $\Theta_{1}=\pi$. (d) $E_{\mathcal{N},j}$ versus $\Theta_{1}$ when $\chi_{j}/\omega_{m}=0.1$. Here $\bar{n}_{j}=10$ and other parameters are the same as those in Fig.~\ref{detuning}.}
\label{N}
\end{figure}

\emph{Entangled optomechanical networks.}---We generalize the DMB approach for optomechanical networks where an optical mode couples to $N\geq 3$ vibrational modes via the optomechanical interactions $H_{\text{opc}}=\sum_{j=1}^{N}g_{j}c^{\dagger}c(d_{j}+d_{j}^{\dagger})$, and the nearest-neighbor vibrational modes are coupled through the phase-dependent phonon-exchange couplings $H_{\text{pec}}=\sum_{j=1}^{N-1}\chi_{j}(e^{i\Theta_{j}}d_{j}^{\dagger}d_{j+1}+\mathrm{H.c.})$ [see Fig.~\ref{Figmodel}(b)]. We have confirmed that these phases are governed by the term $\sum_{\nu =1}^{j-1}\Theta_{\nu}$ ($j\in[2,N]$)~\cite{seeSM}, and hence we assume $\Theta_{1}=\pi$ and $\Theta_{j\in[2,N-1]}=0$ in our simulations.

We demonstrate that when turning off synthetic magnetism (i.e., $\chi_{j}=0$), there exists only a single bright mode $\mathcal{B}=\sum_{j=1}^{N}\delta d_{j}/\sqrt{N}$
and ($N-1$) dark modes, with the $l$th dark mode expressed as $\mathcal{D}_{l\in[1,N-1]}=\sum_{j=1}^{N}\delta d_{j}e^{2\pi i(j-\frac{N+1}{2})l/N}/\sqrt{N}$.
In the presence of synthetic magnetism (i.e., $\chi_{j}\neq0$), all the dark modes are broken by tuning $\Theta_{1}\neq2n\pi$ for an integer $n$~\cite{seeSM}. This provides a possibility of \emph{switching} between the DMB and DMU regimes in optomechanical networks.

We reveal that light and all the vibrations are \emph{separable} ($E_{\mathcal{N},j}=0$, see lower horizontal solid lines) in the DMU regime, but they are \emph{entangled} ($E_{\mathcal{N},j}>0$, see upper symbols) in the DMB regime [see Figs.~\ref{N}(a,b)]. Larger entanglement for optomechanical networks can be achieved for the red-sideband resonance ($\Delta\approx\omega_{m}$) and $\chi_{j}/\omega_{m}\in[0.1,0.15]$ when $\Theta_{1}=\pi$ [see Figs.~\ref{N}(a,b,c)]. Physically, the resulting synthetic gauge fields lead to breaking all the dark modes, and make the light-vibration entanglement networks feasible [see Fig.~\ref{N}(d)]. This indicates that the entanglement networks, with immunity against dark modes, can be realized by applying the DMB mechanism to optomechanical networks. 

\emph{Discussions and conclusions.}---Our scheme can be  implemented using either photonic-crystal optomechanical-cavity configurations~\cite{Fang2017NP} or electromechanical systems~\cite{Ockeloen-Korppi2018Nature,Ockeloen-Korppi2021Science}, where synthetic magnetism can be, respectively, induced by employing two auxiliary-cavity fields or coupling two vibrations to a superconducting charge qubit~\cite{seeSM}. We made detailed parameter analyses and numerical simulations by performing a comparison between the experimental parameters and our simulated parameters~\cite{seeSM}.
For coupled optomechanical networks, a potential experimental challenge concerning the coherent and in-phase pump of the mechanical modes by a common cavity field should be overcomed in realistic experimental setups. Note that the entanglement addressed here (between the intracavity-optical and vibrational modes) is different from that between the output-light and mechanical modes~\cite{Bowen2015,Genes2008PRA3}. The latter entanglement can be analyzed using the input-output relation and the intracavity field-vibration couplings~\cite{Bowen2015,Genes2008PRA3}.

In conclusion, we showed how to achieve both dark-mode-immune and noise-tolerant entanglement via synthetic magnetism. We revealed that both bipartite entanglement and full tripartite inseparability arise from the DMB mechanism, without which they vanish. In particular, our simulations indicated that the threshold phonon number for preserving entanglement could reach \emph{three} orders of magnitude of that in the DMU regime. This study could enable constructing large-scale entanglement networks with dark-mode immunity and noise tolerance.


\begin{acknowledgments}
D.-G.L. thanks Dr. Wei Qin and Dr. Yu-Ran Zhang for valuable discussions. J.-Q.L. was supported in part by National Natural Science Foundation of China (Grants No. 12175061, No. 11822501, No. 11774087, and No. 11935006), the Science and Technology Innovation Program of Hunan Province (Grants No. 2021RC4029 and No. 2020RC4047), and Hunan Science and Technology Plan Project (Grant No. 2017XK2018). A.M. is supported by the Polish National Science Centre (NCN) under the Maestro Grant No. DEC-2019/34/A/ST2/00081. F.N. is supported in part by:
Nippon Telegraph and Telephone Corporation (NTT) Research,
the Japan Science and Technology Agency (JST) [via
the Quantum Leap Flagship Program (Q-LEAP) program,
the Moonshot R\&D Grant Number JPMJMS2061, and
the Centers of Research Excellence in Science and Technology (CREST) Grant No. JPMJCR1676],
the Japan Society for the Promotion of Science (JSPS)
[via the Grants-in-Aid for Scientific Research (KAKENHI) Grant No. JP20H00134],
the Army Research Office (ARO) (Grant No. W911NF-18-1-0358),
the Asian Office of Aerospace Research and Development (AOARD) (via Grant No. FA2386-20-1-4069), and
the Foundational Questions Institute Fund (FQXi) via Grant No. FQXi-IAF19-06.
\end{acknowledgments}

\end{document}